\documentclass[fleqn,usenatbib]{mnras}

\usepackage{newtxtext,newtxmath}

\usepackage[T1]{fontenc}

\DeclareRobustCommand{\VAN}[3]{#2}
\let\VANthebibliography\thebibliography
\def\thebibliography{\DeclareRobustCommand{\VAN}[3]{##3}\VANthebibliography}

\newcommand{\bicho}{LHAASO\,J2108+5157}
\usepackage{graphicx}	
\usepackage{amsmath}	

\title[NIR observations of LHAASO J2108+5157]{Unmasking LHAASO J2108+5157: Near Infrared Insights into a Mysterious TeV Source}

\author[J. Mart\'i et al]{
Josep Mart\'i,$^{1}$\thanks{E-mail: jmarti@ujaen.es (JM)}
Pedro L. Luque-Escamilla,$^{2}$
Josep M. Paredes$^{3}$
and Jos\'e Mart\'inez Aroza$^{4}$
\\
$^{1}$Dept. F\'isica. EPS Ja\'en. Universidad de Ja\'en. Campus Las Lagunillas s/n 23071 Spain\\
$^{2}$Dept. Ingenier\'ia Mec\'anica y Minera, EPS Ja\'en. Universidad de Ja\'en. Campus Las Lagunillas s/n 23071 Spain\\
$^{3}$Dept. F\'isica Qu\`antica i Astrof\'isica, Institut de Ci\`encies del Cosmos, Universitat de Barcelona, IEEC-UB, Mart\'i i Franqu\`es 1, 08028 Barcelona, Spain\\
$^{4}$Dept. Matem\'atica Aplicada, Universitad de Granada, Campus Fuentenueva, 23071 Granada, Spain
}

\date{Accepted XXX. Received YYY; in original form ZZZ}

\pubyear{\the\year{}}

\begin{document}
\label{firstpage}
\pagerange{\pageref{firstpage}--\pageref{lastpage}}
\maketitle

\begin{abstract}
LHAASO J2108+5157 is one of the few ultra-high energy $\gamma$-ray sources in the LHAASO catalogue without secure counterpart at longer wavelengths. Several Galactic scenarios have been proposed, including an evolved supernova remnant and a pulsar wind nebula. Yet, no shocked gas, shell-like structure, or compact pulsar candidate has been identified. Follow-up observations with VERITAS and the LST-1 prototype have not firmly clarified its nature. A recent microquasar candidate from GMRT radio data remains uncertain. Here we present the first dedicated near-infrared study of the field, combining deep $JHK_s$ imaging with narrow band observations targeting the H$_2$ v=1–0 S(1) line. Our observations were initially planned  to encompass the full source region, but now only partially cover the latest updated position and size of LHAASO J2108+5157. We find no evidence of shocked emission, 
extended nebular structures, or an accreting compact object signature in the covered field. The GMRT radio source, despite its jet-like morphology, exhibits near-infrared properties incompatible with both a Galactic microquasar and a nearby radio galaxy, discouraging an association with the gamma-ray emission. Our analysis reveals no convincing counterpart within the positional uncertainty, leaving LHAASO J2108+5157 as an enigmatic ultra-high energy emitter that requires deeper observations.

\end{abstract}




\section{Introduction}

The $\gamma$-ray source \bicho, detected at energies above 100 TeV, remains one of the most enigmatic objects 
in the ultra-high energy (UHE) sky. It was discovered by the LHAASO Collaboration \citep{Cao2021_Discovery} as the only source in its survey without a firmly identified counterpart at longer wavelengths. 
It appears in the first LHAASO catalogue \citep{Cao2024_Catalog} as 1LHAASO\,J2108+5153u . From the published data, the source shows a hard spectrum extending beyond 100 TeV with no evident cutoff, suggesting a possible Galactic PeVatron accelerator \citep{Cao2021_Discovery,Abe2023_Multiwavelength,Adams_2025}. 

In the high-energy (HE) GeV band, the \emph{Fermi}-LAT source 4FGL\,J2108.0+5155 lies at an angular separation of $\sim$0.13°, with photon index and flux in the 1–100 GeV range that challenge a trivial association with the UHE emission \citep{Abe2023_Multiwavelength}. Deep follow-up observations in the
very-high energy (VHE) TeV band by VERITAS ($\sim$40 h), HAWC ($\sim$2400 d), and the LST-1 prototype have yielded only upper limits or weak hints of extended emission. In particular, HAWC detected extended emission in the 3–146 TeV range with a spectral index $\sim$2.45, 
whereas VERITAS observations did not confirm a TeV point source at the LHAASO~J2108+5157 position 
 \citep{Adams_2025}. Similarly, the LST-1 obtained $\sim$49 h of quality data, reporting only a marginal ($\sim$2.2 $\sigma$) excess above 300 GeV, but no clear identification \citep{Abe2023_Multiwavelength}. 
In the X-ray domain, a dedicated 3.8 h \textit{XMM-Newton} observation yielded a non-detection, placing stringent limits on synchrotron emission and disfavouring low–magnetic-field leptonic models \citep{Abe2023_Multiwavelength,Adams_2025}.  
Pre-discovery Swift–XRT observations likewise resulted in non-detections, providing only upper limits \citep{Stroh_2013}.

Several interpretations have been proposed. Galactic scenarios include an evolved supernova remnant (SNR) illuminating a molecular cloud \citep{DeSarkar2023_SNRModel,Mitchell2024_SNRCloud}, or a pulsar wind nebula (PWN) or TeV halo powered by an undetected energetic pulsar \citep{Abe2023_Multiwavelength,Adams_2025}. High-resolution CO observations have identified molecular clouds spatially coincident with the $\gamma$-ray excess, supporting a hadronic origin through $\pi^{0}$-decay from cosmic-ray interactions with ambient gas \citep{deLaFuente2023_MC}. However, no SNR or PWN has been firmly detected in other wavebands, and some authors argue that the GeV emission attributed to 4FGL\,J2108.0+5155 may originate from an unrelated source \citep{Abe2023_Multiwavelength}. Moreover, recent observations with the upgraded
 Giant Metrewave Radio Telescope (GMRT) at 650 MHz revealed a candidate radio source with 
bipolar-jet 
morphology suggestive of a microquasar, though unconfirmed \citep{Mahanta2024_uGMRT}.  

The absence of a clear counterpart at radio, optical, or infrared wavelengths, together with the incomplete viability of standard Galactic scenarios and the unknown distance, makes LHAASO\,J2108+5157 a persistent mystery. Motivated by this, we have undertaken a systematic  near-infrared (NIR) study of the region, combining archival datasets with dedicated observations at the Calar Alto Observatory (CAHA) in Spain. Section~2 describes the observations and the results, and Section~3 discusses implications, concluding that 
there is no convincing observational evidence for a credible accelerator responsible for the gamma-ray emission from \bicho, leaving the origin of this source still unresolved.

\section{NIR observations and results}

As shown in the Introduction, several multi-wavelength studies have targeted the region of LHAASO\,J2108+5157 \citep[e.g.][]{DeSarkar2023_SNRModel,Mitchell2024_SNRCloud,Stroh_2013,Abe2023_Multiwavelength,Adams_2025,deLaFuente2023_MC}, yet none has revealed a convincing low-energy counterpart. To reassess this situation we complemented the existing datasets with dedicated NIR observations using the CAHA 3.5-m telescope, specifically designed to search for faint or diffuse signatures associated with previously proposed SNR and PWN interpretations.

The CAHA campaign was conducted with the OMEGA2000 wide-field camera on 18 June 2024. 
We used the best position information available at that time pointing the telescope to the centroid of the Gaussian model-fit to LHAASO photons, 
as reported in the discovery paper \citep{Cao2021_Discovery}. 
By doing this, the full extent of the updated LHAASO source is now only partially covered in our observations (see Fig.~1). 
For the HAWC, LHAASO KM2A and Fermi-LAT 95\% error regions, this corresponds to a geometric coverage of 46.2\%, 49.2\% and 51.0 \%.
Yet,  we capture 65.8\%, 62.3\% and 51.5\% of the total localization probability based on an integrated 2D Gaussian model, respectively. 
Additionally, NVSS J210803+515255\footnote{\tt https://heasarc.gsfc.nasa.gov/W3Browse/all/nvss.html} is located almost at the edge of the field of view.
 We anticipate here that this fact introduced some difficulties in the analysis of this radio source that, months later, 
was revealed as a possible candidate counterpart as accounted in next Section.

The CAHA run included deep $JHK_s$ images together with narrow-band observations using the NB2.122 and NB2.144 filters, centered respectively on the H$_2$ v = 1--0 S(1) transition (2.1218\,$\mu$m) and its nearby continuum. Their difference provides a sensitive tracer of shocked molecular gas that could arise from energetic outflows or from the interaction of accelerated particles with the ambient medium. 
Standard NIR data reduction was performed with the IRAF software package
involving flat-field, sky-subtraction, median-combining of individual frames and astrometric plate solution on the final images. Accurate astrometry
relied on several hundreds of reference stars retrieved from the {\it Gaia} EDR3 catalogue\footnote{{\tt https://gea.esac.esa.int/archive/}}
and proper motion corrected, yielding typical rms residuals of 0.02-0.03 arc-seconds.
Photometric zero points were finally tied to Two Micron All Sky Survey (2MASS) stars\footnote{{\tt  https://irsa.ipac.caltech.edu/Missions/2mass.html}} in the 
field of view. 

These datasets provide, to our knowledge, the deepest NIR view of the original \bicho\ Gaussian centroid
 to date.
 We estimate that our continuum-subtracted H$_2$ image reaches a $3\sigma$ surface-brightness limit of
$\approx 3 \times 10^{-16}\ \mathrm{erg\ s^{-1}\ cm^{-2}\ arcsec^{-2}}$.
This sensitivity is comparable to that of the UWISH2 survey \citep{Froebrich2011},
which has successfully detected extensive shocked molecular gas associated with
Galactic supernova remnants and jet--ISM interaction regions \citep[e.g.,][]{Lee_2019, Ioannidis_2012}.
The absence of detected H$_2$ features therefore disfavors a significant Galactic shocked-gas counterpart within the observed field. 
.

\begin{figure}
\includegraphics[width=\columnwidth]{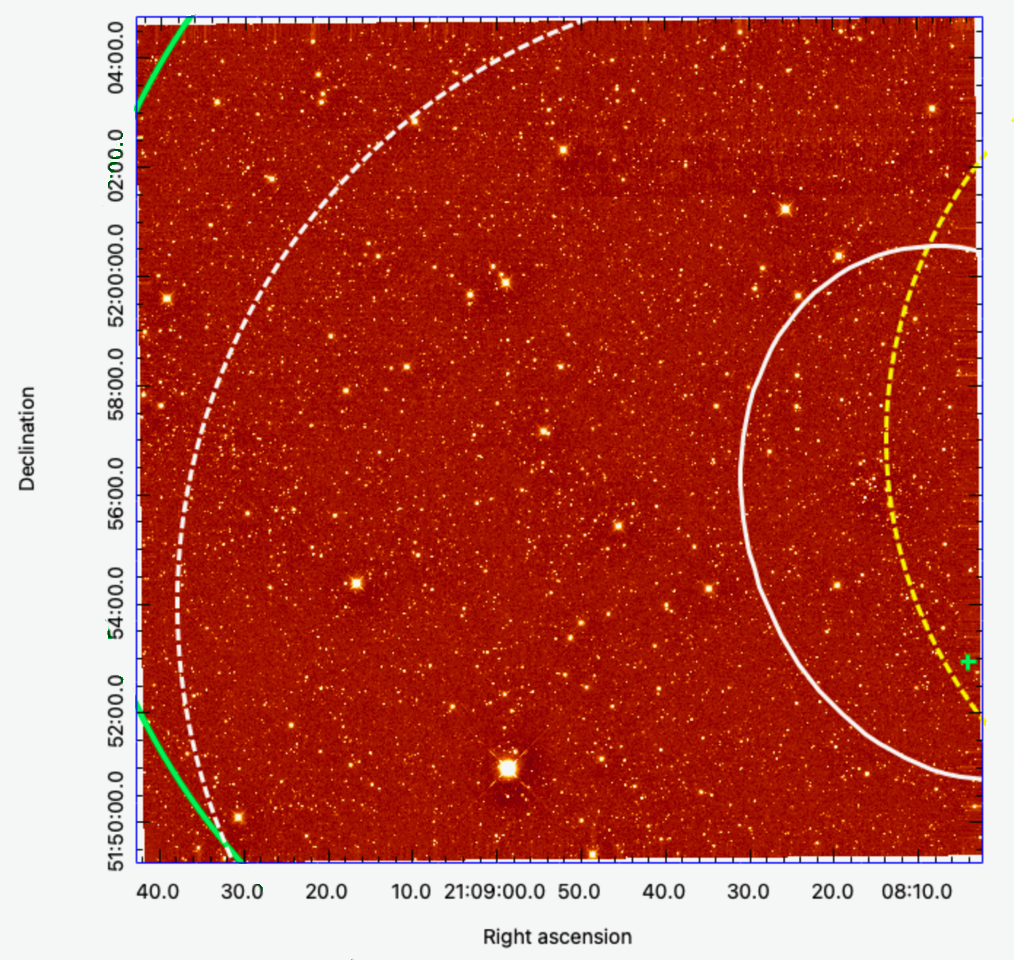}
    \caption{CAHA near-IR field around \bicho\ in $K_s$ filter. Overlaid are the partial confidence regions of the Fermi-LAT (white ellipse), HAWC (green ellipse), and LHAASO (white and yellow dashed ellipses for KM2A and WCDA detectors, respectively) detections. The position of the radio candidate microquasar \citep{Mahanta2024_uGMRT}, consistent in position with all VHE/UHE detections, is marked with a green cross. 
    }
    \label{fig:CAHA}
\end{figure}

As previously stated, we intended  to search for morphological signatures expected from SNRs or PWNs, such as diffuse filaments, arc-like structures, or compact knots of shocked gas. However, no such features were detected in any of the CAHA data. The narrow-band H$_2$ map displayed no excess above the continuum image, and the broad-band frames revealed neither extended nebular emission nor point-like sources with unusual colours. Fig.~\ref{fig:CAHA} displays the resulting $K_s$ image, with the high-energy confidence regions overlaid, that is representative of the entire datasets.

\section{Mining the archives and results}

\begin{figure*}
\includegraphics[width=.9\linewidth]{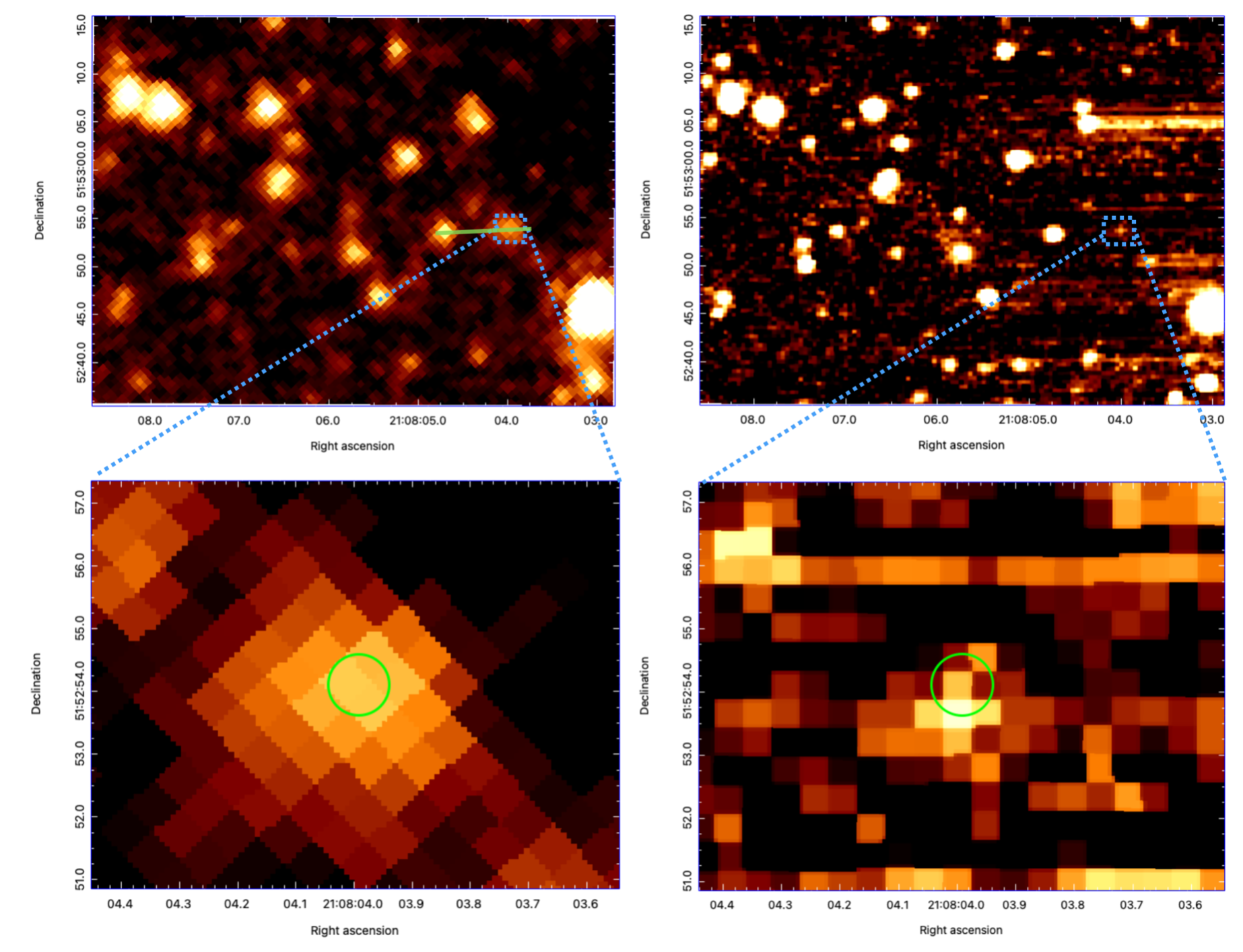}
\caption{
\textit{Left:} \emph{Spitzer}/IRAC\,2 image and a zoom-in of the NIR counterpart of the candidate microquasar proposed by \citet{Mahanta2024_uGMRT}. The green line indicates the path of the brightness profile (slice) used in Fig. 3. The accurate VLASS position of the radio core is marked in the zoomed view with a green circle representing the 95\% confidence radius. 
\textit{Right:} Same as the left, but with the CAHA $K_s$-band image. 
}
\label{fig:OIR}
 \end{figure*}

Given the absence of CAHA-detected structures that could plausibly trace a Galactic remnant, we extended our search to archival datasets to identify any compact object or non-thermal emitter capable of powering the observed HE and UHE radiation. 
We inspected catalogues from radio to X-rays and found no significant source within the overlapping \emph{Fermi}-LAT, HAWC, LST-1, and LHAASO error contours, with a single exception realized later.
The only object of interest emerging from archival data 
is the radio source already reported by \citet{Mahanta2024_uGMRT}, 
also coincident with 
NVSS J210803+5152551\footnote{\tt https://heasarc.gsfc.nasa.gov/W3Browse/all/nvss.html}.
These authors observed with the GMRT, and
detected it at 650\,MHz with a flux density of $27.6 \pm 8.6$\,mJy. Remarkably, their map displays  a well-defined bipolar structure extending over $\sim 24''$. Such a radio morphology led them to suggest that the object may be a previously unknown Galactic microquasar as mentioned above. 
We have additionally examined the ongoing VLA Sky Survey\footnote{{\tt https://science.nrao.edu/vlass}} (VLASS) to assess whether the jet structure is more clearly defined, but the jet appears resolved in the image with no improvement over the GMRT data excluding a sharper view of  the central radio core.
By direct measurement on the VLASS images, this central component is located at standard coordinates
R.A. = 21$^h$ 08$^m$ $03.99 \pm 0.02^s$ and DEC = $+51^{\circ} 52^{\prime} 54.1\pm 0.2^{\prime\prime}$, which provides a much more accurate position than that of GMRT due to the lack of extended emission.

In addition to NVSS, the GMRT jet source is also present in other radio surveys
such as Westerbork Northern Sky Survey\footnote{{\tt https://heasarc.gsfc.nasa.gov/W3Browse/all/wenss.html}} (WENSS) 
and the Canadian Galactic Plane Survey\footnote{{\tt https://www.cadc-ccda.hia-iha.nrc-cnrc.gc.ca/en/cgps/}} (CGPS) .
 The resulting spectral index of $\alpha \approx -0.7$ is fully compatible with non-thermal emission expected from a microquasar. To test this scenario, we inspected our CAHA observations at the source position to search for a possible NIR counterpart. Owing to its critical 
 location near the western edge of the mosaic, only the $K_s$ band provided a minimum coverage. 
 A faint infrared source, with $K_s \simeq 19$\,mag, is certainly 
 detected although at the $4.5 \sigma$ level. 
 Its position is R.A. = 21$^h$ 08$^m$ $04.00 \pm 0.06^s$ and DEC = $+51^{\circ} 52^{\prime} 53.6\pm 0.5^{\prime\prime}$, that agrees with the
 coordinates of the VLASS radio core within astrometric error.
 To complement the CAHA data, we retrieved the \emph{Spitzer}/IRAC images obtained on 12 November 2014 (Program 60020, ObsID 31113). A compact source also coincident with the radio core is clearly detected in IRAC\,1 (3.6\,$\mu$m) and IRAC\,2 (4.5\,$\mu$m). The measured fluxes and magnitudes are summarized in Table~\ref{tab:IR_fluxes}, and the corresponding images are shown in Fig.~\ref{fig:OIR}.
No \textit{Gaia} data is available at this position.

\begin{figure}
\includegraphics[width=\columnwidth]{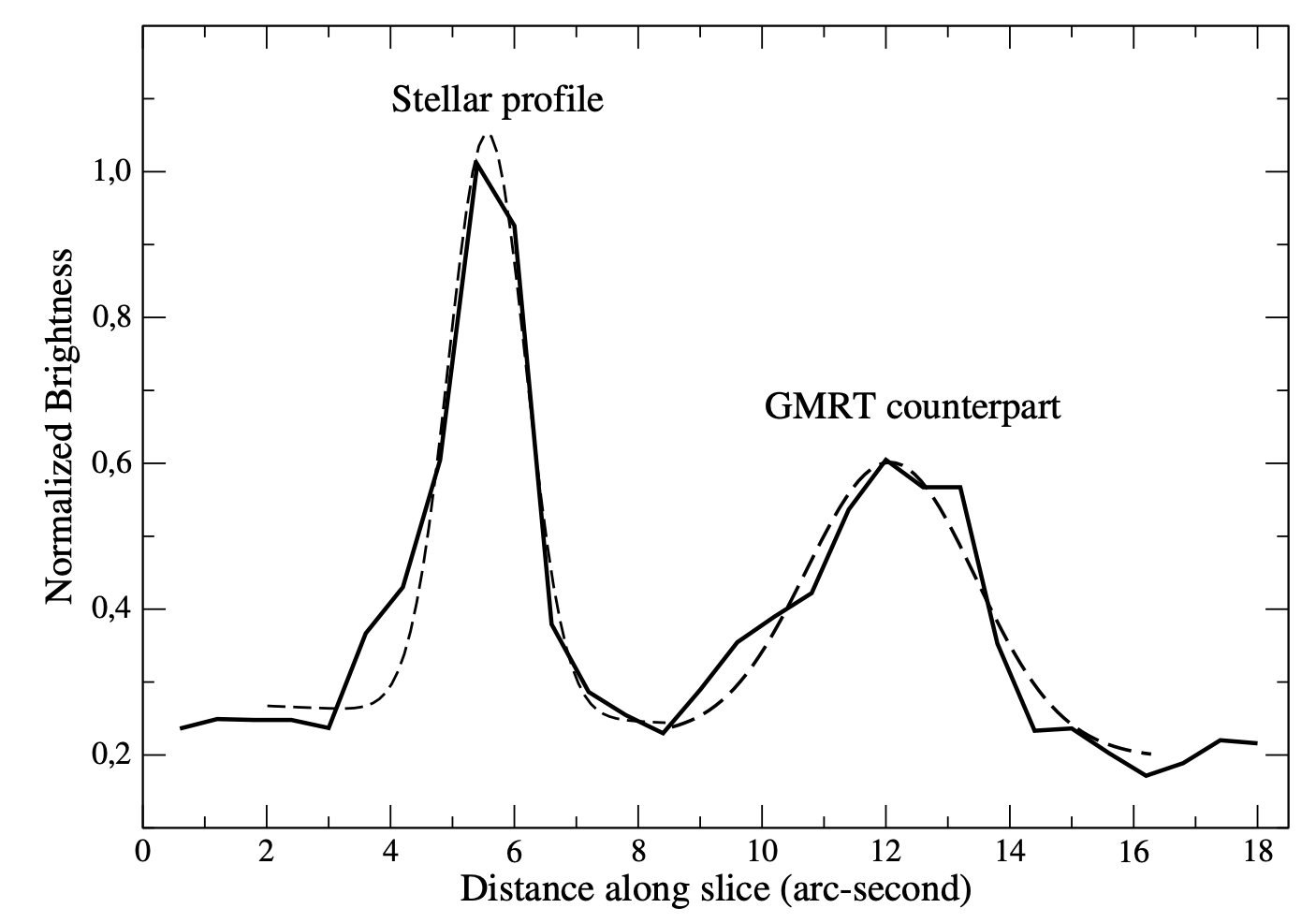}
    \caption{Spitzer IRAC\,2 normalized brightness
profile of the proposed \bicho\ NIR counterpart candidate to the GMRT  radio source, sliced
along the east–west direction so as to pass through the brightness
peaks of both our target and the nearby star located immediately
to the east. The GMRT candidate clearly exhibits an extended morphology.
 Dashed lines show the best-fitting 2D elliptical Gaussian models projected along the cut direction. The source profile is significantly broader than the stellar PSF, indicating that the infrared emission is spatially resolved. The uncertainty of the normalized profile is negligible compared to the observed broadening. 
    }
    \label{fig:morfol}
\end{figure}

To assess whether the infrared emission is resolved, we extracted an East--West brightness profile across the IRAC\,2 image (Fig.~\ref{fig:morfol}), selecting a cut that simultaneously includes the source and a nearby field star. 
The stellar profile is narrow whereas the profile at the source position is visibly broader.
We estimated the spatial extent of the proposed infrared counterpart by fitting 2D elliptical Gaussian models with \textsc{jmfit} task to the IRAC images, allowing the major and minor axes to vary independently.
 The geometric-mean FWHM values are $2.3\pm0.4''$ (IRAC\,1) and $3.2\pm0.7''$ (IRAC\,2), both slightly larger than the empirical point-spread function (PSF) measured from seven nearby field stars ($1.9\pm0.2''$ for both filters), and the reference PSF for these cameras \citep[$\simeq 1.7''$;][]{Fazio_2004}. After quadrature deconvolution, the intrinsic sizes are $1.4\pm0.5''$ and $2.5\pm0.8''$, respectively, indicating that the source is marginally resolved in IRAC\,1 and clearly resolved in IRAC\,2.
This morphology argues against a point-like stellar counterpart and instead points to the presence of extended infrared emission.

\begin{table}
\centering
\caption{Infrared fluxes of the candidate microquasar counterpart.}
\label{tab:IR_fluxes}
\begin{tabular}{lccc}
\hline
Band & $\lambda_{\rm eff}$ [$\mu$m] & Flux density [$\mu$Jy] & Magnitude \\
\hline
$K_s$ (CAHA) & 2.16 & $0.013\pm 0.006$ & $19.0 \pm 0.4$ \\
IRAC\,1 & 3.6 & $0.101  \pm 0.009$ & $16.11  \pm 0.10$ \\
IRAC\,2 & 4.5 & $0.082 \pm 0.008$ & $15.85  \pm 0.10$\\
\hline
\end{tabular}
\end{table}

\section{Discussion and conclusions}
\label{sec:discussion}

Our multiwavelength inspection of the region surrounding LHAASO~J2108+5157, including deep CAHA imaging, 
archival X-ray data, and publicly available radio datasets, reveals no evidence of shocked gas, supernova-remnant-like structures, or any extended nebular emission. The only conspicuous object within the joint LHAASO--Fermi--HAWC error contours is the extended GMRT radio source recently proposed as a microquasar by \citet{Mahanta2024_uGMRT}. Its two-sided jet morphology and its non-thermal spectral index make it the only plausible candidate emerging from our search.
The radio jets reported by those authors span $\sim 24.4''$ tip-to-tip, compatible with parsec-scale jets at Galactic distances typical for microquasars \citep[see, \textit{e.g.},][]{Fender_2006}. Thus, a microquasar origin cannot be dismissed on size alone.

However, the NIR counterpart to the radio core has $K_s \simeq 19$ and is apparently extended. Such a faint magnitude places strong constraints on the nature of any putative donor star. High-mass X-ray binary microquasars typically host O/B companions, which would be far brighter even under heavy extinction. 
While a low-mass X-ray binary scenario remains possible \citep[heavily absorbed low-mass microquasars such as GRS\,$1758-258$ 
near the Galactic Centre still have $K_s \simeq 13.6$, see $e.g.$ ][]{Smith_2025}, the extremely faintness of this object renders such scenario highly unlikely.

Moreover, no X-ray emission is detected at the position of the radio core, despite the fact that microquasars are, by definition, X-ray binaries. 
While heavy extinction could in principle explain this fact, the absorbed luminosity would be reprocessed into the infrared. The extremely faint $K_s \simeq 19$ counterpart exhibits no such excess, effectively ruling out this possibility. This, together with  the extended NIR morphology, further weakens the Galactic interpretation. 
The radio structure itself is strongly reminiscent of classical FR~I jets \citep{FR74} showing hints of deceleration, morphology not typically associated with microquasars. 
These arguments collectivelly point towards an extragalactic nature for the GMRT source.

FR~I radio galaxies are indeed detected at TeV energies 
 \citep[\textit{e.g.}][]{Cao_2024_NGC,Alfaro_2025},
but attenuation by the Extragalactic Background Light (EBL) restricts the transparency of the Universe above $100\,\mathrm{TeV}$ to redshifts $z\lesssim 0.01$. At such distances the implied physical jet length would be unusually small for a typical radiogalaxy ($\lesssim 1\,\mathrm{kpc}$ for $24.4''$), yet the decisive constraint arises from the NIR flux: radio-loud AGN and FR~I hosts are luminous ellipticals with typical absolute magnitudes ${M_K \simeq -25}$ to $-26$ \citep{Dunlop_2003, OlguinIglesias2016}.
Even adopting a conservative ${M_K \simeq -23}$ for an $L^\star$ elliptical host, the observed $K_s$ magnitude would place the source far beyond $z=0.01$. Matching $K_s \simeq 19$ at very low redshift would require an unrealistic faint host (${M_K\gtrsim -14}$), inconsistent with the known population of FR~I population \citep{Ledlow1996, Mingo2019}.
Thus, an extragalactic interpretation compatible with UHE transparency is ruled out.

We therefore conclude that the GMRT source is almost certainly an unrelated background radio galaxy with an intrinsically faint accreting core whose FR~I morphology has led to confusion in previous identifications. Its NIR faintness, inferred host luminosity, and lack of X-ray emission collectively preclude it from being the counterpart to the VHE and UHE emission detected from \bicho.

In summary, we have analyzed the available radio and X-ray catalogues, along with our own NIR and narrow-band filter data from CAHA, to search for evidence of a plausible counterpart to the \bicho\ source and, at least in the region we observed, the only reasonable suspect was the extended bipolar GMRT radio source.
We have astrometrically located its NIR counterpart, which appears as a very faint object with $K_s \simeq 19$ mag and an apparently extended morphology.
However, these properties turn out to be 
inconsistent with both a Galactic microquasar and a very nearby radio galaxy 
under the physical constraints imposed by its multiwavelength properties.
Consequently, none of the objects observed within our partial coverage of \bicho\ can satisfactorily account for the observed gamma-ray emission, leaving the identification of a convincing counterpart uncertain. 
Future deep multiwavelength observations will be crucial to reveal the nature of this enigmatic source, whose true identity remains stubbornly concealed.

\section*{Acknowledgements}

The authors acknowledge support from projects PID2022-136828NB-C41 and PID2022-136828NB-C42 funded by the Spanish MCIN/AEI/10.13039/501100011033 and ``ERDF A way of making Europe'',
and
through the Unit of Excellence María de Maeztu 
awards to the Institute of Cosmos Sciences (CEX2024-
001451-M).
We also acknowledge support from project AST22\_0001\_16 with funding from NextGenerationEU funds; the “Ministerio de Ciencia, Innovaci\'on y Universidades” and its Plan de Recuperaci\'on, Transformaci\'on y Resiliencia” and “Consejer\'{\i}a de Universidad, 
Investigaci\'on e Innovaci\'on of the regional government of Andaluc\'{\i}a.
Based on observations collected at the Centro Astron\'omico Hispano en Andaluc\'{\i}a (CAHA) at Calar Alto, 
proposal 24A-3.5-003, operated jointly by Junta de Andaluc\'{\i}a and Consejo Superior de Investigaciones Cient\'{\i}ficas (IAA-CSIC).

\section*{Data Availability}

The data used in the preparation of this Letter are public and can be accessed in their respective databases.
Reduced CAHA images are available on request to the authors.



\bibliographystyle{mnras}
\bibliography{LHAASO_J2108+5153} 





\bsp	
\label{lastpage}
\end{document}